\documentclass[fleqn,twoside]{article}
\usepackage{espcrc2}
\usepackage{epsfig}

\title{Inflation}

\author{I.I.~Tkachev\address[MCSD]{CERN Theory  Division,
        CH-1211 Geneva 23, Switzerland.}}%

\def\ltap{\raisebox{-.55ex}{\rlap{$\sim$}} \raisebox{.4ex}{$<$}}

\def\lsim{\mathrel{\ltap}}

\begin{document}

\begin{abstract}
  Inflationary cosmology is reviewed. Particular attention is given to
  processes of creation.
\end{abstract}

\maketitle

\section{Introduction}

A typical universe should have had Plankian size, live Plankian time and
contain 1 particle. 
Yet, the observable Universe contains $10^{90}$ particles in it and had
survived $10^{65}$ Plankian times. Where does it all came from ?  In other
words, why is the Universe so big, flat and old ($t > 10^{10}$ years),
homogeneous and isotropic ($\delta T/T \sim 10^{-5}$), why does it contain so
much entropy and does not contain unwanted relics?  These puzzles of classical
cosmology were solved with invention of Inflation \cite{Guth:1980zm}.

\subsection{Getting something for nothing}

Stress-energy tensor $T^{\mu\nu}$ which drives the expansion of 
a homogeneous universe
can be characterised by two parameters, energy density, $T^0_0 = \rho$,
and pressure $T^i_j = -p\delta^i_j$.
Conservation of energy and momentum 
{$T^{\mu\nu}_{\,\,\,\,\,\,\,\,\,;\nu}=0$}
in an expanding Friedmann universe takes a simple form
$
{\dot{\rho}} +3H({\rho}+{p})=0 \, . 
$
Here $H$ is a Hubble parameter, $H \equiv \dot{a}/a$,
and $a(t)$ is a scale factor which describes the expansion of the Universe.
Consider the stress-energy tensor {$T_{\mu\nu}$} of a vacuum. Vacuum has
to be Lorentz invariant, hence {$T_{\mu}^{\,\nu}=V\,
\delta_{\mu}^{\,\nu}\;$} and we find {$\;p =-$}{$\rho$}.  Therefore,
the energy of a vacuum stays constant despite the expansion.  In this
way, room for matter full of energy could have being created.  
It remains to find out how vacuum energy was converted into radiation 
at some later stage.

\subsection{Horizon problem and the solution}

The size of a causally connected region (horizon) scales in proportion to
time, $R_{\rm H} \propto t$. On the other hand, the physical size of a
given patch grows in proportion to the scale factor, $R_{\rm P} \propto
\;a(t) \propto t^\gamma$.  Exponent $\gamma$ depends upon equation of
state, $\gamma = 1/2$ for radiation and $\gamma = 2/3$ for matter
dominated expansion. In any case, for the ``classical'' Friedmann Universe 
$\gamma < 1$ and horizon expands faster than volume. 
Take the largest visible patch today. It follows that
in the past it should have contained many casually disconnected
regions and the question arises why the Universe is so homogeneous at
large scales ? This problem can be solved if during some period of
time the volume had expanded faster than the horizon. During such a period
the whole visible Universe can be inflated from one (``small'')
causally connected region.  Clearly, this happens if $\gamma > 1$, 
which means $\ddot{a} > 0$.  Either of these two conditions can be used
as definition of an inflationary regime.

Using the Einstein equation $\ddot{a} = - {4\pi}G a (\rho +
3p)/{3}$ we find that the inflationary stage is realized when 
$p < - \rho /3$. If $p = - \rho$ we have de Sitter metric
and the Universe expands exponentially. 

A crucial and testable prediction of Inflationary cosmology is a flat
Universe, $\Omega =1$ (as usual, $\Omega (t)$ is the ratio of current
and critical energy densities, $\Omega (t) \equiv 8\pi G\rho/3H^2$).
Indeed, Einstein equations can be cast into the 
form ${\dot{a}^2}(\Omega(t) -1) = {\dot{a}^2_0}(\Omega_0 -1)$. Accelerated
expansion, $\ddot{a} > 0$, increases $\dot{a}$ and therefore
drives $\Omega(t)$ to 1.

The first model of inflation was de-facto suggested in
\cite{Gurovich:xg}. De Sitter expansion appeared as a result
of vacuum polarization effects in a one-loop order -- too complicated
to be sure that higher order corrections are unimportant.

\subsection{Arranging for a vacuum}

Consider {$T_{\mu\nu}$} of a real scalar field {$\phi$}
$$
T_{\mu\nu}=\partial_{\mu}\phi\, \partial_{\nu}\phi - 
g_{\mu\nu}\, {\cal L}
$$ 
with the Lagrangian 
${\cal L}=\frac{1}{2}\partial_{\mu}\phi\, \partial^{\mu}\phi - 
V(\phi)$. 
In a state when all derivatives of {$\phi$} are negligible, 
$\partial_{\mu}\phi \simeq 0$, the stress-energy 
tensor of a scalar field is that of a vacuum, 
{$\; T_{\mu\nu} \simeq V(\phi)\,g_{\mu\nu}\;$}.

There are two basic ways to arrange 
{$\phi \simeq$ const} 
and hence to imitate the {vacuum}-like state.

{1.} The simplest possibility was suggested by {A. Guth} in his original
paper \cite{Guth:1980zm}. 
Consider potential $V(\phi)$ which has a local minimum with
a non-zero energy density separated from the true ground state by a
potential barrier.  The Universe will be trapped in the meta-stable minimum
for a while and expansion will diminish all field gradients. Then the Universe
enters a vacuum state.  Subsequent phase transition into the true minimum
ends inflation and creates the radiation phase.
Today the model of Guth and its variants based on potential barriers is 
good for illustration purposes only. It did not stand up to
observations since inhomogeneities which are created during the
phase transition into the radiation phase are too large \cite{Guth:uk}.

2. A. Linde was first to realise that things work in the simplest possible
set up \cite{Linde:gd}.  Consider potential
\begin{equation}
V(\phi)=\frac{1}{2}m_\phi^2\phi^2 \; .
\label{chaotic}
\end{equation}
Equation of the field motion in expanding Universe is
$
\ddot{\phi}+3H\dot{\phi}+m_\phi^2\phi=0 \; .
$
If {$H \gg m $} the ``friction'' is too big and the field
(almost) does not move. Therefore time derivatives in $\; T_{\mu\nu}$
can be neglected and inflation starts (in sufficiently homogeneous
patch of the Universe). Hubble parameter in this case is given by
{ $H \approx m {\phi}/{M_{\rm Pl}} $} and we see that inflation
starts if initial field value happen to satisfy ${\phi > M_{\rm
Pl}}$. During inflationary stage the field slowly rolls down the potential
hill. This motion is very important in the theory of structure creation.
Inflation ends when ${\phi \sim M_{\rm Pl}}$. At this time
field oscillations start around potential minimum and latter decay
into radiation. In this way matter was likely created in our Universe.

\section{Unified theory of creation}

Small fluctuations of any field obey
\begin{equation}
\ddot{U}_k ~+~ [k^2 + m^2_{\rm eff} (\tau)]\; U_k = 0 \, .
\label{ModEq}
\end{equation}
Effective mass is time dependent here because of the expansion of the
Universe. Because $m_{\rm eff}$ is time dependent, it is not possible
to keep fluctuations in vacuum.  If one arranges to put oscillators
with momentum $k$ into the vacuum, they will not be in vacuum at a latter time
since this vacuum would correspond to wrong value of the field mass.  

Some remarks are in order:
\begin{itemize}
\item Eq.~(\ref{ModEq}) is valid for all particle species      
\item Equation looks that simple in conformal reference frame
        $ds^2 = a(\tau)^2\;(d\tau^2-dx^2)$. Everywhere in this chapter
        a ``dot'' means derivative with respect to $\tau$.
\item Of particular interest are ripples of space-time itself
\begin{itemize}
\item curvature fluctuations (scalar fluctuations of the metric)
\item gravitons (tensor fluctuations of the metric)
\end{itemize}
\item $m_{\rm eff}$ may be non-zero even for massless fields
\begin{itemize}
\item graviton is the simplest example \cite{Grishchuk:1974ny},
        $m^2_{\rm eff} = - {\ddot{a}}/{a}$ 
\item $m^2_{\rm eff}$ for curvature fluctuations has similar structure
  \cite{lukash} with $a$ being replaced by $a\dot{\phi}/H$ 
\end{itemize}
\item For conformally coupled, but massive scalar
        $m_{\rm eff} = m_0 \, a(\tau)$ 
\end{itemize}
Creation was only possible because nature is not conformally-invariant.
Otherwise $m_{\rm eff} =0$ and vacuum remains vacuum forever.

\subsection{Sources of creation}

Amplitudes ${U}_k$ in Eq. (\ref{ModEq}) are quantum operators and
a theory of creation reduces to the theory of Bogolyubov
transformations or to a theory of particle creation in homogeneous
time varying classical background. There are two important instances
of such background in cosmology:
\begin{itemize}
\item Expansion of space-time, $a(\tau )$
\item Motion of the inflaton field, $\phi (\tau )$ 
\end{itemize}
Both can be operational at any epoch  of creation
\begin{itemize}
\item During inflation 
\item While the inflaton oscillates (reheating)
\end{itemize}
During inflation superhorizon size perturbations of metric are
created which give seeds for Large Scale Structure (LSS) formation and
eventually lead to formation of galaxies, the Solar system and all the
rest which we can see around us. During reheating matter itself is
created. Overall there are four different situations (two sources
times two epochs). If coupling to the inflaton is not essential, the
corresponding process will be called ``pure gravitational creation''
in what follows.  Let me consider all four possibilities 
in turn, starting from

\subsection{Gravitational creation of metric perturbations \cite{GCMP}}

During inflation the motion of the inflaton field is slow, while the expansion
of the Universe is fast. It follows that relevant cosmological scales
encompass small $\Delta \phi$ interval.  E.g. in the model Eq. (\ref{chaotic})
the whole visible Universe is inflated away while the inflaton field $\phi$
changes from $4M_{\rm Pl}$ to $M_{\rm Pl}$. Potential $V(\phi)$ should be
relatively ``flat'' over this range of $\Delta \phi$ to maintain the
inflationary regime.  We may conclude that observables should essentially
depend on a first few derivatives of $V$ and the shape of the potential
outside this region of $\Delta \phi$ is irrelevant.  One may construct
dimensionless quantities {(slow roll parameters)} out of potential derivatives
\begin{eqnarray}
&& \epsilon \equiv \frac{M_{\rm Pl}^2}{16\pi}\left(\frac{V'}{V}\right)^2\, 
, \nonumber\\
&& \eta \equiv \frac{M_{\rm Pl}^2}{8\pi}\frac{V''}{V}\, .  
\label{SRPar}
\end{eqnarray}
The value of potential itself during this period is also
relevant and defines the value of the Hubble parameter, $H^2(\phi) =
8\pi G V(\phi )/{3}$.

Solutions of Eq. (\ref{ModEq}) with vacuum initial conditions 
give for the power spectra of scalar (curvature) and tensor
(gravity waves) perturbations 
\begin{eqnarray}
&& P(k)_{S} =  \frac{1}{\pi\epsilon }\frac{H^2}{M_{\rm Pl}^2}\, 
, \nonumber\\
&& P(k)_{T} =  \frac{16}{\pi}\frac{H^2}{M_{\rm Pl}^2}\, .
\label{STspectra} 
\end{eqnarray}
These spectra can be approximated as power law functions
\begin{eqnarray}
&& P(k)_{S} =  P(k_0)_{S} \left(\frac{k}{k_0}\right)^{n -1}\, 
, \nonumber \\
&& P(k)_{T} =  P(k_0)_{T} \left(\frac{k}{k_0}\right)^{n_T}\, . 
\label{nSnTdef} 
\end{eqnarray}
where scale dependence $k$ enters via weak dependence
of $H$ on the current field value (current means here 
corresponding to the moment when
the scale gets bigger than the horizon and evolution of the mode $k$
freezes out).  One expects spectra to be nearly scale invariant since
the field (almost) does not move. Indeed, expanding in slow roll
parameters one finds 
\begin{eqnarray}
&& n - 1 =  2\eta - 6 \epsilon\, 
, \nonumber\\
&& n_T = - 2 \epsilon\, . 
\label{nSnTslowroll} 
\end{eqnarray}

\subsubsection{Consistency relation}

According to Eqs. (\ref{STspectra}) the ratio of power in tensor to scalar
perturbations is equal to $16\epsilon$. On the other hand the exponent
of tensor perturbations is also proportional to $\epsilon$, see 
Eq. (\ref{nSnTslowroll}). This gives ``consistency relation'',
$
n_T \simeq -0.14\, r \,  ,
$
where the ratio of tensor to scalar power is expressed through the
ratio of directly measurable respective contributions to quadrupole
CMBR anisotropy, $r \equiv {C^{T}}/{C^{S}}$. Verification of the
consistency relation should give major peace of evidence that
the Inflation did happened.

\subsubsection{Testing Inflation}

Predictions of inflationary theory can be tested measuring
CMBR anisotropy and power spectra of galaxies distribution.  All tests
completed so far are in agreement with predictions.  Latest CMBR data
\cite{CMBRA} imply that the Universe is flat $\Omega_0 = 1.02 \pm
0.04$. The spectrum of perturbations is nearly scale invariant, $n_s =
0.97 \pm 0.1$, and Gaussian \cite{Wu:2001wu}. 
Overall normalisation of CMBR spectrum fixes inflaton parameters. 
In the simplest case of Eq. (\ref{chaotic}) $m_\phi \simeq 10^{13}$ GeV.

Consistency relation was not tested yet
since it requires measurement of tensor perturbations.  In forthcoming
CMBR experiments 
tensor and scalar modes may be disentangled. This can be done unambiguously
because tensor and scalar perturbations contribute differently to the
polarisation of CMBR (for a review see \cite{Hu:1997hv}) which hopefully will
be measured.

Measured position of the first acoustic peak in CMBR not only tells us that
the Universe is flat, but also that the isocurvature perturbations are ruled
out as the primary block of structure formation. While during inflation both
types of perturbations, curvature and isocurvature, can be produced, the
models which do not involve inflation (e.g. network of cosmic strings) produce
isocurvature fluctuations.  Therefore this is an important test of inflation
\cite{AcPeaksTD}.  In principle, curvature perturbations may be mimicked by
causal processes \cite{Turok:1996wa}, but polarisation measurements will also
give unambiguous proof that density perturbations are of superhorazon origin
\cite{Spergel:1997vq}.

\begin{figure}
\vspace{-.2cm}
\hspace{-.6cm}\epsfig{file=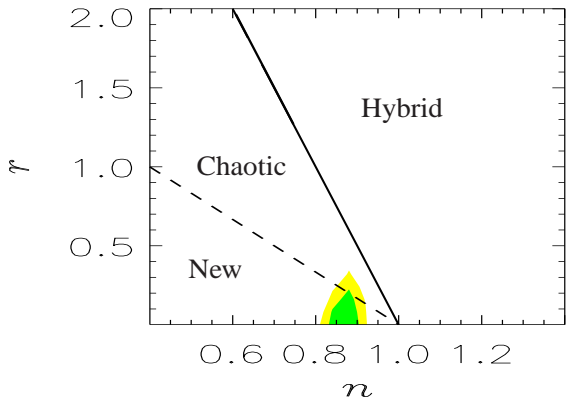,height=6.cm,width=8cm,%
bbllx=90,bblly=445,%
bburx=520,bbury=780,%
clip=}
\vspace{-1cm}
\caption{Different models of inflation predict $n$ and $r$ to be found 
in specified regions. Shaded area indicates parameter range favoured 
by CMBR and LSS measurements.}
\label{fig:regionsANDdata}
\end{figure}

While the tensor mode is not measured yet, it is restricted by CMBR anisotropy
measurements. Different models of inflation occupy well defined and different
regions in the (r,n) parameter plane \cite{Dodelson:1997hr}, see
Eqs.~(\ref{SRPar}), (\ref{nSnTslowroll}).  Regions, specified for new
\cite{new}, chaotic \cite{Linde:gd} and hybrid \cite{Linde:1993cn}
inflationary models are shown in Fig.~\ref{fig:regionsANDdata}.  Other
inflationary models are also bounded to these regions.  For example,
``natural'' inflation \cite{Freese:1990rb} fits the region of ``new''
inflationary model. Parameter space favoured by current CMBR and LSS
measurements is also shown \cite{Kinney:2001nc} as a shaded area (note however
that size and shape of this region depends upon priors used). E.g., for the $
V \propto \phi^p $ chaotic inflation model all this means that {$p < 6$} at
99\% c.l.  \cite{Kinney:2001nc}, while the hybrid model of inflation is
disfavoured.

\subsection{Gravitational creation of matter}

Let us consider the gravitational creation of matter. The source of creation
is $ m_{\rm eff} = m_0 \,a(\tau)$. While matter is created in tiny amounts by
this process, it is most effective for heaviest particles.  If such superheavy
particles do exist in nature, this process naturally leads to Superheavy Dark
Matter (SDM) \cite{Chung:1999zb,Kuzmin:1998uv}. If these particles are
unstable but long lived, they can explain \cite{Berezinsky:1997hy} puzzling
Ultra-High Energy Cosmic Ray (UHECR) events.

\subsubsection{Friedmann cosmology}

It is particle mass which couples quantum fluctuations to the background
expansion and serves as the source of particle creation. Therefore we expect
for the abundance of Super-Heavy particles
\begin{equation}
n_{\rm SH} \propto m_{\rm SH}^3 a^{-3}\, . 
\label{nXFr}
\end{equation} 
It is the expansion of the Universe which causes particle creation, therefore
creation is most efficient at time $\tau_0$ when $H \approx m_{\rm SH}$, while
(comoving) particle number is an adiabatic invariant at later times. We
expect that coefficient in Eq. (\ref{nXFr}) should not be much smaller unity
if we normalize $a(\tau_0) = 1$.  It follows that stable particles with
{$m_{\rm SH} > 10^9$ GeV} will overclose the Universe if there was Friedmann
singularity in the past -- Friedmann cosmology and SDM mutually exclude each
other \cite{Kuzmin:1998uv}.

\subsubsection{Inflationary cosmology}

In inflationary cosmology there is no singularity and the Hubble constant
is limited, $H \lsim m_\phi$. Therefore, the production of particles with 
$m_{\rm SH} > H \sim 10^{13}$ GeV is suppressed.

\begin{figure}
\hspace{-0.5cm}\includegraphics[width=8cm]{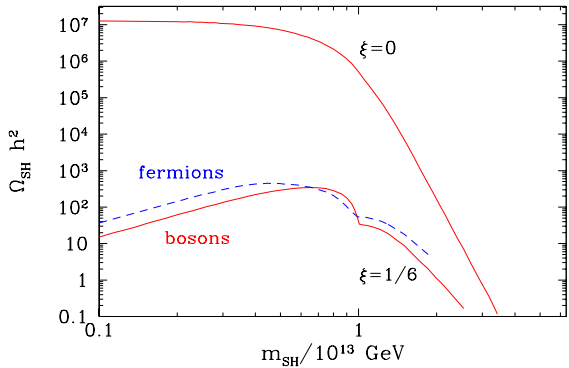} 
\vspace{-1cm}
\caption{Present day density of gravitationally created Superheavy Dark
  Matter. Solid lines correspond to minimally or conformally coupled bosons,
  dotted line describes abundance of fermions.}
\label{fig:GravCr}
\end{figure}

The present day ratio of the energy density in $SH$-particles to 
the critical energy density is shown in Fig.~\ref{fig:GravCr}
\cite{Kuzmin:1998uv} (see also \cite{Chung:1999zb}).
Super-Heavy particles with the mass $\rm few \times 10^{13}\; GeV$
are excellent candidates for SDM and progenitors of UHECR. 
If it will be proven that UHECR are due to decays of
SDM, it will mean that Friedmann expansion was 
preceded by some other epoch, likely by Inflation.

Consistency with observations requires $\Omega_{\rm SH} h^2 \lsim 0.3$.  We
see that light particles, $m_{\rm SH} \ll 10^{13}$ GeV, are overproduced by
many orders of magnitude, unless their coupling to curvature is conformal $\xi
= 1/6$.  This poses a serious danger \cite{Kallosh:1999jj} since
supresymmetry and suprgravity models predict many such particles, e.g.
moduli, gravitino, etc. They may be dangerous relics,
even if unstable, since they can easily survive to post-nucleosinthesis
epoch and their decay products destroy $^4$He and D nuclei by 
photodissociation \cite{Ellis:1982yb}.

\subsection{Decay of the inflaton oscillations}

While bosons and fermions are created equally efficiently by a pure
gravitational mechanism, coupling to inflaton uncovers deep
differences between them. Effective mass of a scalar $X$ 
(interaction Lagrangian $~L_{\rm int}=\frac{1}{2} g^2 \phi^2 X^2$) and
of a fermion $\psi$ (interaction Lagrangian $~L_{\rm int} = g \phi
\bar{\psi}\psi$) is given by the following expressions
\begin{eqnarray}
{\rm scalar} X:&& \;\; m^2_{\rm eff} = m_X^2 + g^2\phi^2(t)\,  
\label{mX} \\
{\rm ~fermion~} \psi:&& \;\; m_{\rm eff} = m_\psi + g\phi(t)\, . 
\label{mF}
\end{eqnarray}
Effective mass of a scalar $X$ depends quadratically upon the 
inflaton field strength and therefore it is always larger
than the bare mass $m_X$. In the case of fermions inflaton field
strength enters linearly and effective mass can cross zero. Even
superheavy fermions can be created easily during these moments of zero
crossing \cite{Giudice:1999fb}.  
(It is easier to create a light field and it is effective mass
which counts at creation. At the end of the day it is bare mass which
counts.)

Coupling $g$ by itself is not relevant for the process of creation,
$g$ always comes in combination with inflaton field strength. To make
dimensionless combination out of it we have to re-scale $g\phi$ by a
typical time scale relevant for creation.  In the present case this
will be period of inflaton oscillations or inverse inflaton mass
\begin{equation} 
g^2 \rightarrow  q \equiv \frac{g^2\phi^2}{4m_\phi^2}.
\end{equation}
Parameter q determines the strength of particle production caused
by the oscillations of the inflaton field. It can be very large 
\cite{Kofman:1994rk} even
when $g$ is small since
${\phi^2}/{m_\phi^2} \approx 
10^{12}$.

\subsubsection{Matter creation: Bose versus Fermi}

Bose-stimulation aids the process of creation of bosons.  Occupation numbers
grow exponentially with time, $n = e^{\mu t}$, which results in a fast,
explosive decay of inflaton
\cite{Kofman:1994rk,Shtanov:1995ce} and creates large classical
fluctuations of all Bose-fields involved. This can have a number of observable
consequences: non-thermal phase transitions
\cite{Kofman:1996fi,Tkachev:1996md}, generation of a stochastic background of
the gravitational waves \cite{Khlebnikov:1997di}, and a possibility for a novel
mechanism of baryogenesis \cite{Kolb:1996jt} are some examples.  

Fortunately, the system in this regime of particle creation became
classical and can be studied on a lattice 
\cite{Khlebnikov:1996mc,Prokopec:1996rr,Felder:2000hq}.

On the other hand in the case of fermions $n \le 1$ at all times because of
the Pauli blocking. This may create an impression that the fermionic channel
of inflaton decay is not important. This is a false impression
\cite{Greene:1998nh}.  Production of fermions can be more efficient compared
to bosons \cite{Giudice:1999fb}.  Vanishing effective mass allows for a larger
Fermi-momentum of created particles and fermions can outnumber bosons.

\begin{figure}
\hspace{-.3cm}\includegraphics[width=8cm]{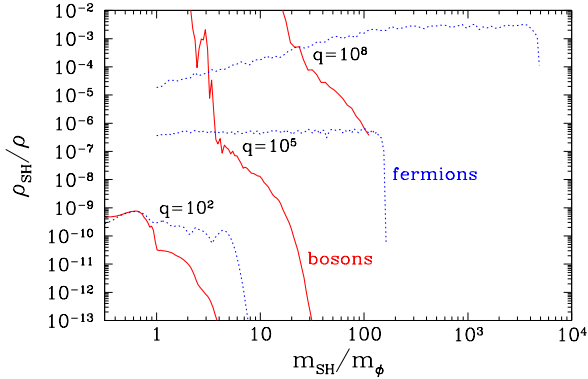} 
\vspace{-1.2cm}
\caption{Effectiveness of Super-Heavy particle production.
Solid lines production of fermions, dotted 
lines production of bosons.}
\label{fig:BvF}
\end{figure}

A fraction of the initial energy density which goes to bosons and
fermions respectively is plotted in Fig.~\ref{fig:BvF} as a function
of particle mass for several values of $q$. Indeed, superheavy
fermions are more efficiently created compared to bosons at the same 
value of $q$.

\subsubsection{Non-thermal phase transitions}

\begin{figure}
\hspace{-.3cm}\includegraphics[width=8cm]{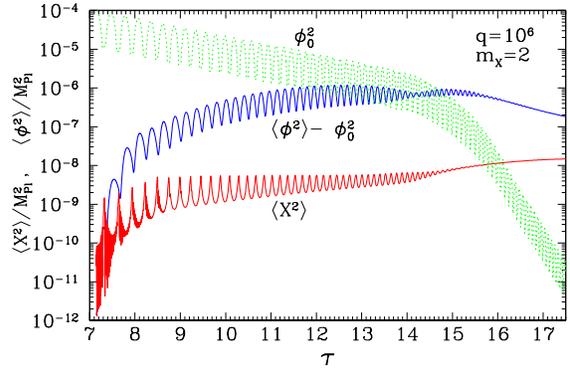} 
\vspace{-1.2cm}
\caption{Late-time dependence of field variances and of inflaton zero mode
$\phi_0$. Time and $m_X$ are in units of $m_\phi$. At $\tau=0$ variances are
in vacuum.}
\label{fig:Variances}
\end{figure}

Bosons of moderate or small mass (compared to inflaton), and sufficiently
large coupling ($q > 10^4$ or $g^2 > 10^{-8}$ \cite{Khlebnikov:1996wr}) are
produced explosively, their density grows exponentially.  Field variances,
$\langle X^2 \rangle$, can reach large values, larger than in a thermal
equilibrium at the same energy density.  These were calculated for different
values of parameters in Ref.~\cite{Khlebnikov:1996mc}. For one choice of $m_X$
and $q$ the time dependence of field variances is shown in
Fig.~\ref{fig:Variances}. For smaller $m_X$ and $q$ variances can reach larger
values. By the final moment of time shown in this figure inflaton zero mode had
already decayed, but the system is still very far from thermal equilibrium.

In the theory of phase transitions the effective mass of the Higgs field in a
medium is
$m_{\rm H}^2 = -\mu^2 + h^2 \langle X^2  \rangle  $.  
If $\langle X^2  \rangle$ is sufficiently large a phase transition can occur.
During preheating in the process of inflaton decay this can have
especially interesting consequences.
Symmetry restoration and subsequent breaking with possible formation of 
topological defects \cite{Tkachev:1998dc}
can happen even in GUTs with large symmetry breaking scale.

\subsection{Creation of matter during Inflation}

While inflaton during Inflation moves slowly, it does move. This motion may
lead to the creation of particles coupled to it. Most effectively and with
observable consequences this occurs in the case of superheavy fermions.
Features in the power spectrum of perturbations which may leave trace in CMBR
and LSS appear at scales corresponding to zero crossings of effective mass.
This can be a probe of Sub-Plankian particle content.  
Multiplet of $N$ particles with~ $m_\psi
\sim M_{\rm Pl}$ and coupling $g > 0.2 / N^{2/5}$ is detectable
\cite{Chung:2000ve}.

\section{Conclusions}

Inflationary cosmology is a beatuful theory. It inputs unknown and (largely)
arbitrary initial conditions and replaces them with testable predictions. Even
the simplest inflationary model, Eq. (\ref{chaotic}), reproduces all relevant
features of observable Universe after adjustment of a single parameter,
$m_\phi \simeq 10^{13}$ GeV.

\end{document}